**Artificial Intelligence Ethics Education in Cybersecurity: Challenges and Opportunities: a focus group report**

Diane Jackson

Doctoral Student in Communication

Sorin Adam Matei

Professor of Communication

Elisa Bertino

Professor of Computer Science

Purdue University

100 University St.

West Lafayette, IN

Contact: Sorin Adam Matei, smatei@purdue.edu

The research reported in this paper was supported by the NSF EAGER award no. 2114680: SaTC-EDU: A Life-Cycle Approach for Artificial Intelligence-Based Cybersecurity Education

Abstract

The emergence of AI tools in cybersecurity creates many opportunities and uncertainties. A focus group with advanced graduate students in cybersecurity revealed the potential depth and breadth of the challenges and opportunities. The salient issues are access to open source or free tools, documentation, curricular diversity, and clear articulation of ethical principles for AI cybersecurity education. Confronting the "black box" mentality in AI cybersecurity work is also of the greatest importance, doubled by deeper and prior education in foundational AI work. Systems thinking and effective communication were considered relevant areas of educational improvement. Future AI educators and practitioners need to address these issues by implementing rigorous technical training curricula, clear documentation, and frameworks for ethically monitoring AI combined with critical and system's thinking and communication skills.

Developments in artificial intelligence have impacted many research and practical areas, including cybersecurity. With new tools and possibilities came new concerns, including a loss of transparency, control over the decision-making process, and the diminishing capability of AI users to judge their ethical implications (Anderson & Rainie, 2022). To assuage these concerns, an area of research investigating ethical AI practices has been burgeoning (d'Aquin et al., 2018; Hagendorff, 2019; Jobin et al., 2019; McLennan et al., 2020; Middelstadt, 2019; Morley et al., 2021; Rességuier & Rodrigues, 2020; Taddeo & Floridi, 2018). However, curricular offerings demanded by specific situations or integrating ethical principles into AI research practices cannot be the only response to these concerns. The next and necessary step should be fostering educational approaches to building a more complex curricular offering connected with critical thinking and ethical AI practices. This would be even more consequential as it involves the downstream impacts of equipping current and future AI practitioners with the opportunity to apply ethical frameworks to the deployment of AI.

In this paper, we investigate the needs and perceptions of those at the receiving end of the learning process, advanced graduate cybersecurity students exposed to AI training and work. We report the findings of a focus group at a large US state university about the technical educational, and ethical challenges that advanced doctoral computer science students have encountered in their educational and professional experiences. Because AI practices and applications vary based on the computer science context and specialty, we have narrowed our scope to the context of cybersecurity education that intersects with AI tools and practices.

Due to the mounting pressures of companies to protect themselves against cyberattacks and to protect the data and privacy of their users, customers, and employees, cybersecurity is an incredibly important area that benefits from the deployment of artificial intelligence while requiring an ethical framework to shape the way that practitioners apply artificial intelligence in the cybersecurity sector. As such, this paper aims to contribute insights into the technical and educational needs faced by individuals who have worked as both advanced graduate computer science students in the cybersecurity sphere.

This paper will begin by reviewing existing literature about the ethical and educational concerns that have accompanied the use of AI in cybersecurity and computer science more broadly. The paper will then discuss how the existing literature led to the primary themes that shaped focus group discussion. We continue by presenting the themes that emerged during the focus group, including discussions about the technical and educational challenges. The paper concludes by discussing how AI ethical principles, critical and systems thinking, and engaged communication can address some issues that emerged during the focus group discussions.

**AI Ethics in Cybersecurity**

Researchers of AI ethics have identified multiple issues in teaching ethics in AI. For instance, there is the issue of developing practical applications from theoretical principles (McLennan et al., 2020; Mittelstadt, 2019; Rainie et al., 2021; Whittlestone et al., 2019). Further, because of the evolving nature of technologies and human needs, it is possible that ethical decisions made today could have unexpected non-ethical outcomes tomorrow (see Morley et al., 2020). However, literature about AI ethics education in the context of cybersecurity issues have predominantly focused on younger students (see Bendechache et al., 2021; Grover et al., 2023; Kilhoffer et al., 2023; Walsh et al., 2023) or on the methodological of technical approaches to deal with AI and computing ethics (Slavkovik, 2020). We must pay more attention to the educational process by which advanced learners professional and research aspirations are prepared to deal with ethical issues in cybersecurity and AI.

Their needs and perceptions have evolved in the context of an exponential growth in cybersecurity research that intersects with AI tools and their ethical implications. As machine learning algorithms and other forms of artificial intelligence have been more widely adopted, many educational or practical concerns have arisen from implementing artificial intelligence in computer science. For instance, most experts agreed that by 2035, systems will not be designed for humans to control most automated decision-making (Anderson & Rainie, 2023). Further, 60% of Americans have reported feeling uncomfortable with their medical provider relying on AI (Tyson et al., 2023) and even more Americans have reported feeling that AI would majorly impact jobholders in the next 20 years (Rainie et al., 2023). To address these concerns, an area of research has emerged focusing on the application of ethical principles to AI use (see d'Aquin et al., 2018; Hagendorff, 2019; Jobin et al., 2019; McLennan et al., 2020; Middelstadt, 2019; Morley et al., 2021; Rességuier & Rodrigues, 2020; Taddeo & Floridi, 2018). Similarly, literature has emerged considering the role of ethics in the cybersecurity space (see Brey, 2007; Christen et al., 2020; Formosa et al., 2021; Himma, 2008; Manjikian, 2018). However, less attention has been directed to the employment of AI in cybersecurity (see Timmers, 2019), and more attention needs to be paid to the ethical practices of engaging AI in the context of cybersecurity (Grover et al., 2023; Richards et al., 2020; Sinha & Huraimel, 2021).

According to Timmers (2019), AI is usually engaged in cybersecurity when AI is used as an element of protecting the security and safety of critical systems. AI has also been used in automated cybersecurity or in weaponizing AI for cyberattacks (see Brundage et al., 2018). Among these uses, concerns about cyberattack or counterattack software using AI have become increasingly important (Brundage, 2018; Taddeo, 2018). The lack of a widely practiced ethical framework that can be implemented into AI, particularly in the context of cybersecurity, makes the risk of weaponizing AI or committing attacks that leverage the weaknesses of AI particularly daunting. Further, AI has complicated the ability to secure networks because it bring about more vulnerabilities that can be hard to identify and address (Michael et al., 2023; Sinha & Huraimel, 2021). However, AI can also be employed as a tool for recognizing threats to cybersecurity (Sinha & Huraimel, 2021). While computer scientists are navigating the paradoxical nature of the relationship between cybersecurity and AI, calls to establish and instill ethical considerations as fixtures in algorithms (Badhwar, 2021) is critical to effective cyber defense (Liu & Murphy, 2020). At the same time, the focus on tools must be doubled by training the next generation of cyber experts to handle the multiple challenges, ethical or otherwise, brought about by the introduction of AI in cybersecurity work.

Regardless of approach, technical or human, ethical use of AI in cybersecurity demands specific criteria and frameworks for defining what is ethical and how to behave or design ethically in the AI space. While many researchers have identified some of the inadequacies with the current AI technology and with specific algorithms (Morley et al., 2020) and have made calls for specific ethical frameworks to be incorporated into AI practice, one recent meta-analysis shows that about 80% of these guidelines prescribe minimal ethical requirements like accountability, privacy, and fairness (Hagendorff, 2020). Further, themes of justice, doing no harm, responsibility, and privacy emerged in over half of the articles analyzed about ethical artificial intelligence (Jobin et al., 2019; Morley et al., 2020). Together, these findings contribute to overarching ethical principles that center around respect to people and environment (beneficence), robustness and security (non-maleficence), fairness (justice), and the accountability and understandability of the AI (explicability) (Jobin et al., 2019; Morley et al., 2020).

However, identifying the conceptual elements of an ethical AI framework is only one aspect of developing ethical AI practices. The true challenge is practically developing, refining, and instilling these ethical guidelines into the fabric of of higher education in cybersecurity, computer science and AI (d'Aquin et al., 2018; McNamara et al., 2018; Morley et al., 2020; Raji et al., 2021). This report focuses on the educational considerations of distilling ethical principles into computer science and cyber security education by reporting findings from a focus group of advanced computer science / cybersecurity students. We structured the research protocol around the themes identified above, especially the core values related to ethical work and design. In addition, we asked exploratory questions about the technical and critical thinking education that should undergird research and professional development in AI driven cybersecurity education. The protocol aimed to reveal existing educational and practical experiences and felt needs in a cybersecurity context.

**Focus Group Protocol and Recruitment**

The focus group was organized at a large US university with strong computer science and cybersecurity research areas. Four students were selected by a call for volunteers from a cybersecurity course dedicated to socio-technical issues, including ethics. The course was developed in the context of an NSF-funded project, which called for creating a course prototype that combines ethical with technical education. The students volunteered to participate and were not compensated. The discussion was carried after the grading period. The conversations were transcribed verbatim. Themes of concern were identified qualitatively, reading across the answers to and discussions around the questions proposed by the protocol. The themes were organized as clusters of comments structured around the questions. The summary below synthesizes the themes and discusses their main ideas while illustrating them with representative quotations.

**Educational Challenges of Teaching AI Ethics in Cybersecurity and Core Ethical Principles**

AI training demands both technical prowess and an ethical understanding of how making choices about the means and ends of the process and the tools involved in cybersecurity might affect the end users or various incidental populations. The focus group probed questions about the participants' understanding of ethics and ethical principles, their presence or absence in coursework, and their level of preparedness for AI-enabled cyber work that might entail ethical choices.

The participants' opinions were divided in terms of what cybersecurity ethics should be concerned with. One participant had a data-centric, privacy, and consent view of ethics. "For me, what comes to mind, is that data has to be taken with permission." (Participant 3) Another focused on fairness: "For example, your data should have information representing an entire population… For example, if you have voice data and the voices are taken from males but there are no female voices, that's just an extreme example where the data themselves are biased and the model will be biased, too." (Participant 1) A third was mostly concerned about explainability "I think AI ethics would be explainability. What the AI thinks should be somehow explained in some way other than what the black box shows." (Participant 3) Trust was the most important ethical concern for a third participant "I think it's not just about the data, it's also about the model itself. Imagine that even if you are given this data in a correct way, the design of the model itself could be prone to errors of how [I] make a decision… So, the decision-making could be ethical… Some models are more prone to bias than others. So, I would like to see how the model is making the decisions. What are the specific math behind it and what makes it different from other models." (Participant 2). This was a concern shared by Participant 4, in a more nuanced way "Let's say we need AI power to do something then can we trust that outcome? That's into data, models, and so on."

However, being concerned about ethics and behaving ethically or designing ethical procedures in AI-based cybersecurity are two different things. Three of the participants noted that they had no formal training in cybersecurity or AI ethics of any kind. Yet, the participants offered specific and meaningful perspectives on what ethical training might look like, mostly extracted from their practical work. Participant 2 recalls the importance of interdisciplinary teaching and of the multidisciplinary preparedness of the instructors: "In my undergrad [studies] there were some requirements, and the professor was actually a lawyer in the law school and there was so much behind that… We had to have a lawyer to tell us about GDPR and how this works and that professor was working at the intersection of computer science and law." (Participant 2)

The participants thought that practice-focused education, organized around specific, real-world effects of cybersecurity work, would be quite useful "For computer science ethics, when we find issues, we have to notify people within 90 days before we report it to the public… Maybe doing a training on that would be helpful." (Participant 3). Overall, the participants agreed that offering a mandatory course or workshop would be beneficial to educating computer science students about AI ethics practically.

However, beyond existing preparedness, generous advice, and felt needs, there was an evident lack of vision and structure in ethical education in cybersecurity.

**Pedagogical / Curricular Concerns Now and in the Future**

The emergence of AI tools and the increasing sophistication of cybersecurity threats compels cybersecurity educators to reconsider how and what they teach. The challenges go beyond educational issues, including technical, social, and ethical concerns. According to the results of our focus group, the biggest educational challenges that computer science students seeking advanced degrees face are a result of tensions between course and research demands, limited course offerings, out-of-date assignments, and disconnection between theory taught in the classroom and practice.

For these students, attention should be placed more on the practice. Prioritizing research responsibilities while managing the workload of course commitments is a challenging practice. The students placed greater value on expanding course offerings to include more specializations in cybersecurity and updating assignments to incorporate newer technologies that have been developed. As such, they emphasized the value they derived from a curriculum that related theoretical principles to practical applications. Professors who can integrate research with coursework and who can incorporate newer technological innovations into students' curriculum and assignments will be able to alleviate these challenges for students.

**Technical Issues**

Cybersecurity enhanced by AI tools demands deep and well-considered educational materials based on a common understanding of the tools, their documentation, use, and prevention methodologies. However, the field is so new and fragmented and dominated by proprietary, commercial systems that are hard to access and understand. The insufficient documentation and lack of an agreed-upon coding language make these issues too complex to simplify for a teaching environment.

The focus group protocol included questions about the students' technical challenges in the coursework. Specifically, we asked about difficulties inherent in the tools and in the educational program.

One of the most important challenges in cybersecurity education focusing on AI topics is the lack of documentation and background information about the most current tools and challenges, an issue echoed by existing research (Timmers, 2019). This shortcoming limits the students basic understanding of what is possible and what they are expected to learn in certain situations. "Systems are black boxes, it's really hard to understand how we can have access to these systems. Part of the process for the last 2-3 years is you have to understand the phone [as an example of a closed boxed system] and how to have access before starting the research on the phone" (Participant 3). "Sometimes, [documentation] is even nonexistent, so you have to try to understand on your own what is happening." (Participant 2)

A second, equally important challenge, is the lack of a standard coding language for cybersecurity projects. While it is expected that researchers might use different approaches rooted in specific coding languages, the lack of a core, "go-to" language creates systems that are different in their very conceptual form and, because of this, in their implementation. For example, Participant 2 observed, "In security, there's not an applied, widely adopted language in which we specify or adopt things, so there are different heterogeneous systems. When people try to put them together, they usually come up with vulnerable, proprietary solutions and this is one of the most critical reasons why we have so many vulnerabilities today because everyone is trying to do their own thing." (Participant 2)

The third problem is the complexity of cybersecurity problems, which can take many forms and present multiple forms of attack that are hard to characterize to put into a specific technical framework. "When you want to investigate what's the cause of the attack, it's not because of a single failure or single problem that exists in the system. Sometimes, the attack combines different weaknesses. Sometimes, the attack involves lots of different disciplines in computer science and sometimes human factors are huge factors in that as well. Sometimes there are insider attacks that cannot be completely prevented by technology." (Participant 1)

There might be multiple points of failure, including human factors residing inside an organization. Conceptualizing all these issues as one integrated framework becomes even more difficult when AI tools are involved, which can mimic or hide human failures. Overall, the technical issues associated with education in AI-enhanced cybersecurity create uncertainties in terms of "what" to learn, with what tools, in what context and with what effect.

**Learning Challenges**

The focus group participants agreed that significant learning obstacles confront AI Cybersecurity education. They mentioned that their programs did not offer sufficient flexibility. The organization of their curricula is disjointed. A disconnect exists between the needs of master's students and the needs of doctoral students. "Depending on the department, the workload is too much… The unified program of master's and Ph.D. may be a problem because the Masters may want more of a certain cybersecurity course whereas the Ph.D. may want more focus on research and less workload." (Participant 2)

Participants also mentioned that in some of the programs they attended, there was a lack of cybersecurity-specific courses. Faculty members only offered a general cybersecurity course without providing students with opportunities to dive more deeply into more up-to-date cybersecurity issues and topics. Participants also mentioned that the assignments that they are tasked with have tended to be outdated exercises irrelevant to the most recent technological innovations. They also mentioned that there can be a disconnect in the course content between the theoretical foundations they are taught and how they apply practically.

One participant gave the example of an assignment he was instructed to complete involved defending against path overflows. However, the participant had recognized that recent defense mechanisms have made path overflows harder to break, which diminishes the utility of this sort of assignment. Participants felt a tension between wanting to study what they will face in the real world and lacking the time to test these harder-to-break, more recent defense mechanisms that have emerged more recently. The student who reflected on the path overflow assignment example concluded that at least some new things currently deployed in systems could be taught to current students.

The overall educational challenge that the focus group reported was the disconnect between theory and practice. Yet, two of the international students who were part of the focus group expressed an appreciation that they have been able to practice what they study while studying at universities in the United States. They reflected on how meaningful this is by stating, "Doing in practice really helps you to understand the issue that you are trying to solve and how to solve it." Participant 4.

**AI tool-specific educational concerns**

The focus group participants remarked that educating the next generation of cybersecurity professionals depends on curricular offerings and up-to-date toolkits that are easy to understand and use. They also pointed to the fact that the lack of AI model transparency limits learning even though recent research initiatives are being developed to explain and visualize model features. On the other hand, participants mentioned the struggle they experienced with understanding which datasets are most compatible with the models and the lack of documentation for both models and datasets. "You need to actually have the background of what model works best with your data. That's actually one of the extra challenges. Even if you have a perfect tool, but you don't know which one you should use." (Participant 1)  Participant 1 also observed  that "Compared to past few years, the libraries are much easier and more efficient to use. There are still problems, though […] it would be better if the tools would add more transparency in their model. Now, it is like you call the function and it acts like a black box and you don't know if it's the correct output. There is some research in AI that is trying to explain the features and how the model is using the features and visualizations that show that."

The presence of hardware-specific code compounds tool-dependent learning issues. - "The dependency on hardware is increasing exponentially. If we try to learn a large model, you need a lot of hardware to power it because it's not cheap." (Participant 3)

The problem is not without hope, considered the participants. They see more and more datasets and toolkits developed for or released for educational purposes. Repositories of AI/ML datasets compatible with cybersecurity work are also increasing in number, and the students would like to see more of them. More important than creating repositories, it is, however, to release datasets that are compatible with the coursework. There should be no disconnect between what the professor teaches and what the student can experiment with. "The availability of data is a big problem for courses. First of all, maybe we should actually encourage the release of datasets for educational purposes only. One of the funding processes that we have at the [local AI repository] provides all kinds of data for network-related research, so I think we should encourage things like this. Another thing with data is that we have some research in machine learning that proposes some model that we can learn surprisingly well from a limited amount of data. Maybe when we design a course, we should also mention that as a point for students who are interested in those topics." (Participant 1)

The technical issues associated with AI training never come alone. They are embedded in the general preparedness of the students in the broader theoretical and methodological areas of statistics, data science, or mathematics. Participants specifically mentioned that having experience with advanced mathematics, statistics, and more advanced programming languages gave them confidence that they would be able to engage AI tools effectively in their careers. Many participants also recognized that they expected to engage AI at some point in their careers, even if they had not yet done so. "I haven't worked with AI much, but I think that I have been given the background to be able to approach a problem that involves AI. If you think about CS, for us, we have an easier way, but if you think outside of computer science, most people have no idea how AI actually works. As a student in CS and getting a Ph.D., I have a nice background and I have easier access to AI even if I haven't used it." (Participant 4)

Cybersecurity encompasses many professional and theoretical aspects, and its practitioners are of many kinds. The participants recognized that those that come from a CS background are naturally privileged, "For us, with a lot of math and stats and a computer language background, it's not hard to get into AI. I've mostly been in systems research, but I have a background in machine learning—mostly mathematical and statistical—so when learning with the tools of the programming language, it's easier for us because we usually work on a lower level of abstraction, like harder levels of language like CC++." (Participant 3) This observation asks what type of cybersecurity AI education should be provided to learners and professionals who are not as technical, especially the front-line, customer-facing ones. Future AI in cybersecurity educational reform should keep in mind the sheer complexity of the problem and the high degree of technical sophistication demanded by the models and their datasets. In the candid words of Participant 1, "I am actually for my grad course taking pure machine learning courses and once you start studying them and diving into them—before it's like you only see the tip of the iceberg, and when you start studying them, even for a very simple model it actually involves lots of math details… it's actually very complicated than you think… if you try the model from scratch, it's like hundreds of lines of code."

Regarding solutions, the participants opined that offering more advanced courses in machine learning earlier at the undergraduate level would be beneficial while acknowledging that doing so would come at the cost of lowering the opportunities to take more general education courses across disciplines and focusing earlier on in computer science-related study. o

"I think from my understanding, the best thing you need to do is look into the black box… they should at least be knowing what's going on inside of implementing a model from scratch to understand how the model is doing what it should do… that level of abstraction is needed for students to understand how they should develop tools. Otherwise, they will be fixed in a box and they will always be missing that." (Participant 3)

On the other hand, Participant 4 observed, "It's a big tradeoff between what do you teach and what do you not teach because if you want to introduce machine learning in undergrad, then you have to sacrifice many other courses. I studied chemistry that maybe I didn't need because I never use chemistry, so perhaps they can restructure some of the focus."

**Broader educational preparedness for work in AI Cybersecurity**

*Systems thinking*

Future-oriented cybersecurity workers trained in AI tools need to be abstract thinkers. They cannot conceptualize their problems in a simplistic, step-wise manner. They need to think in terms of systems and systems-thinking. The focus group protocol included a set of questions about systems thinking, including

- When I say "system thinking," what thoughts come to your mind? What does it mean to you?

- How important is systems thinking for using AI in cybersecurity work?

- Can you give an example, be specific?

- Thinking about your own coursework, how well prepared are you to use system thinking in your studies?

- What type of training and education do you think you need to improve your preparedness?

Some of the participants defined systems thinking as paying attention to the more minute details. In general, they believed to be prepared to use systems thinking. They suggested that teaching system thinking is to teach around open problems that do not have a current solution, which forces the students to find new connections between disparate details and think for themselves. Systems thinking encapsulates a form of critical thinking.

One participant also noted that although thinking about the sum-total of a problem, in cybersecurity, the specific steps within a process are also necessary "It is not just how you feed the data to the model, you have to do some processing. So, all of these steps in the way could cause problems or it could cause changing something in between that might either give you something that's wrong or might give you something that's highly efficient. So, it's a pipeline." (Participant 2) This observation did not mean to discount the importance of system thinking. It did, however, point to the fact that in cybersecurity systems are procedural and sequential. You cannot easily jump from one part of the system to another.

Overall, all participants agreed that systems thinking should play a larger role in educating students about AI and cybersecurity: "In computer science, everything needs to be systems thinking. AI, maybe more than anything else. We work with systems, so everything we think about needs to be logical, cause-effect. It's necessary." (Participant 4) More importantly, students felt qualified, even at this stage of their learning and careers, to engage in Systems Thinking "We do try to understand how things can go wrong and what is the reason behind those problems." (Participant 2)

The participants also offered their pedagogical suggestions for specific ways in which students might be able to practice Systems Thinking. Participant 3 suggested that professors include more open problems that do not currently have solutions to train students how to engage in systems thinking.

**Communication skills in cybersecurity and ethics**

Cybersecurity professionals interact with many types of stakeholders throughout the entire chain of information technology use. They require superior ability to communicate clearly and effectively about the technical or social problems and solutions they encounter. The focus group participants repeatedly highlighted that cybersecurity professionals must communicate extremely well with non-experts, be they internal or external. Technical prowess may be stymied if the worker cannot communicate effectively about his or her actions. Participant 2 shared that "[Communication] is extremely important, and it should be shaped based on who the audience is… One needs to have a skillset to communicate what the research is, what the problem is, what the audience is as well." Participant 4 put it in more direct terms "The most difficult thing is to explain to my friends and family what I do. If I have to talk to my sister and tell her what I do, it's very complicated and… it's too much to teach."

Given the significance of the topic, the participants have offered numerous insights into what can be done to improve communication skills in cybersecurity, especially when dealing with complex issues such as AI and ethical issues. They contributions focused one a central theme: talk to everyone as if they had no expertise: "The most difficult thing is to explain to my friends and family what I do. If I have to talk to my sister and tell her what I do, it's very complicated and… it's too much to teach." (Participant 4) "We needed to change a lot of what we had to say if only the managers and people with less CS background. If you want to convey the impact of their products, it's different than with technical experts." (Participant 2) "The most important thing is imagining if you talk to someone who doesn't have any background with your work and try to explain that in three minutes and not losing that person is the rule." (Participant 1).

Suggestions for improved teaching and training methods abounded, from classes to workshops and internships. Participant 4 suggested "[a] class that focuses on [communication] that gets people from outside of [the computer science] environment [as critical audience]. Because it's true that we have to show a paper or present a paper, we still have an audience who has a lot of technical background, so it was still a discussion that's very technical. But, if you get out of that, it wouldn't be as easy." Participant 1 offered "communication workshops like talking communication arts or 3-Minute Thesis competition would be good practice."  For internships, Participant 3 proposed that the organizers and sponsors "specifically ask you to make your presentation in such a way that doesn't have that many technical details so that listeners don't get lost… it's a work-in-progress, btu I am still learning. There are always pros and cons. I want to talk about more technical stuff but some folks don't want to hear more technical stuff, it's a delicate balance." (Participant 3)

**Conclusion**

Cybersecurity research, practice, and education need to adapt to the constant changes in the world of computing. The emergence of AI-drive computing challenges researchers, practitioners, educators, and students to rethink their questions, procedures, curricula, and learning styles. The present study, albeit limited in footprint, reveals that graduate students are justifiably concerned about the pace and timeliness of the education they need to keep up with the latest AI-driven developments in cybersecurity. They raised legitimate issues about the profound disconnect between education and practice. A core issue during the conversation was the difficulty of studying emerging threats and solutions in real time. The main culprit is the closed-box nature of many of the systems they study. This demands new and universally available toolkits that can replicate and even anticipate cybersecurity issues.

The focus group also highlighted the lack of formal ethical training and clear ethical criteria for educating the next generation of cybersecurity experts charged with using AI tools. This can be mitigated through specialized courses in AI ethics and a broader, critical thinking education at all levels. In this respect, systems thinking is essential for training students to confront the complex nature of AI-based cybersecurity threats critically and creatively. The focus group finally revealed the role of effective communication in translating cybersecurity threats and solutions, together with their ethical implications in non-expert language.

The present results offer only a limited but directive view on the main challenges and opportunities confronting cybersecurity education. We hope that further research, using more complex instruments and larger samples will offer a more diverse and in-depth perspective on this emergent issue.